\documentclass[journal,11pt,onecolumn]{IEEEtran}
\ifCLASSINFOpdf
\else
\fi

\usepackage[utf8]{inputenc} 
\usepackage[T1]{fontenc}
\usepackage{url}
\usepackage{ifthen}
\usepackage{cite}
\usepackage{amsmath}

\usepackage{todonotes}
\usepackage{graphicx}
\usepackage{amssymb}
\usepackage{amsthm}
\usepackage{amsfonts}
\usepackage{mathtools}
\usepackage{dsfont}
\mathtoolsset{showonlyrefs=true}
\usepackage{algorithm}
\usepackage[noend]{algpseudocode}

\usepackage{color}
\usepackage{float}

\newcommand{\SNR}{{\rm SNR}}
\newcommand{\PAM}{{\rm PAM}}
\newcommand{\SK}{{\rm SK}}
\newcommand{\ZSK}{{\rm ZSK}}
\newcommand{\zoom}{{\rm zoom}}
\newcommand{\iter}{{\rm iter}}
\newcommand{\target}{{\rm target}}
\newcommand{\dB}{{\rm dB}}

\newcommand{\rem}[1]{}

\newtheorem{lemma}{Lemma}
\newtheorem{corollary}{Corollary}
\newtheorem{theorem}{Theorem}

\newcommand{\bre}{\begin{equation}}
\newcommand{\ere}{\end{equation}}

\newcommand{\ee}\]
\newcommand{\bra}{\begin{eqnarray}}
\newcommand{\era}{\end{eqnarray}}
\newcommand{\bfg}{\begin{figure}[hbtp]}
\newcommand{\efg}{\end{figure}}
\newcommand{\bver}{\begin{verbatim}}
\newcommand{\ever}{\end{verbatim}}
\newcommand{\bit}{\begin{itemize}}
\newcommand{\eit}{\end{itemize}}
\newcommand{\ben}{\begin{enumerate}}
\newcommand{\een}{\end{enumerate}}

\newcommand{\half}{\frac{1}{2}}

\newcommand{\bsigma}{\boldsymbol\sigma}

\newcommand{\Exp}{\mbox{E}}

\newcommand{\bK}{{\bf K}}

\newcommand{\Var}{\mathrm{Var}}
\newcommand{\Cov}{{\mathrm{Cov}}}

\newcommand{\E}{{\mathrm E}}

\newcommand{\baa}{\begin{eqnarray*}}
\newcommand{\eaa}{\end{eqnarray*}}

\newcommand{\bM}{{\bf M}}

\newcommand{\cE}{{\cal E}}

\newcommand{\cN}{{\mathcal{N}}}

\newcommand{\beginproof}{\noindent \textbf{Proof: }  }
\newcommand{\finproof}{\noindent $\Box$\\}
\newcommand{\defined}{\triangleq}

\def\argmin{\mathop{\rm argmin}}

\def\defined{\: {\stackrel{\scriptscriptstyle \Delta}{=}} \: }

\newfont{\boldlarge}{msbm10 scaled 1100}

\newcommand{\comment}[1]{}

\newlength{\tmpbigbar}

\begin{document}
%
\title{Feedback Channel Communication with Low Precision Arithmetic}
%
%
%

\author{Yonatan Urman and David~Burshtein
	\thanks{This research was supported by the Israel Science Foundation (grant no. 1868/18).}
	\thanks{Y.\ Urman is with the school of Electrical Engineering, Tel-Aviv University, Tel-Aviv 6997801, Israel (email: yonatanurman@mail.tau.ac.il).}
	\thanks{D.\ Burshtein is with the school of Electrical Engineering, Tel-Aviv University, Tel-Aviv 6997801, Israel (email: burstyn@eng.tau.ac.il).}
}
\maketitle

\begin{abstract}
The problem of communicating over an additive white Gaussian noise channel with feedback, using low precision arithmetic, is considered. The Schalkwijk-Kailath (SK) scheme is known to achieve an error probability that decays double exponentially in the number of interaction rounds, for any rate below channel capacity.
However, SK is also known to suffer from numerical issues.
Transmission close to channel capacity requires a moderate number of interaction rounds. This may lead to a huge constellation size. Furthermore, the internal variables of the scheme decay to zero exponentially fast. As a result, the SK scheme fails when implemented with low precision variables, which are widely used in hardware implementations.
In this work we propose a new, modified scheme termed Zoom-in SK (ZSK), which breaks the SK protocol into several stages. Each stage comprises several SK iterations followed by a synchronized zoom step. The zoom-in allows the receiver and transmitter to keep the scheme's parameters relatively large such that low precision arithmetic can be used even for a large rate or a large number of interaction rounds. We prove that the new scheme achieves approximately the same error probability as SK while not suffering from numerical issues. We further verify our results in simulation and compare ZSK to the original SK scheme.
\end{abstract}


%
\IEEEpeerreviewmaketitle

\section{Introduction} \label{sec:Introduction}
\begin{figure}[H]
	\centering
	\includegraphics[width=0.9\columnwidth]{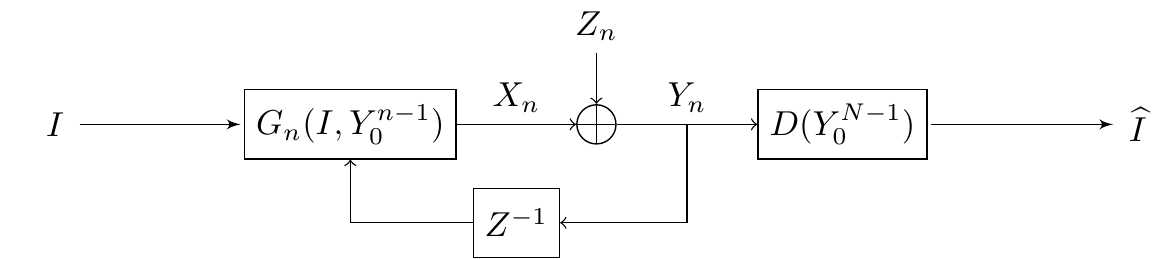} 
	\caption{Noiseless AWGNC with feedback. The element $Z^{-1}$ represents a unit delay.}
	\label{fig:feedback_cha}
\end{figure}

Consider the discrete memoryless additive white Gaussian noise channel (AWGNC) with noiseless feedback, shown in Figure \ref{fig:feedback_cha}. The transmitter attempts to transmit a message, $I \in \{0,1,\ldots,M-1\}$, using $N$ channel uses, where at each time step it transmits over the AWGNC, and then it receives noiseless feedback from the receiver. More precisely, at each time step, $n=0,1,\ldots,N-1$, the transmitter sends $X_n = G_n(I, Y_0^{n-1})$ over the AWGNC. We assume an average power constraint $P$ at the input to the channel. The receiver attempts to decode the transmitted message, $I$, using $\widehat{I} = D(Y_0^{N-1})$. The communication rate in bits per channel use is $R = \log_2 M / N$.
While it is well known that feedback cannot improve the capacity of point to point communications \cite{cover_book}, there exist schemes where it can significantly reduce complexity and / or improve reliability (reduce error probability). For the case of AWGNC with feedback, the Schalkwijk-Kailath (SK) scheme \cite{SK_1,SK_2,SK_3} can achieve any rate below the channel capacity with error probability that decreases double exponentially in the number of interaction rounds (iterations), $N$.
In \cite{shayevitz2011optimal} a generalized approach for feedback communication, using posterior matching, was presented. Special cases of posterior matching are the Horstein scheme for the binary symmetric channel \cite{horstein1963sequential} and the SK scheme for the AWGNC. 

The SK scheme conveys its message $I$ to the receiver using an $M$-PAM constellation. In the beginning, the transmitter transmits an $M$-PAM symbol representing the message $I$ over the AWGNC. Then, in the following iterations, it transmits an error correction signal to the receiver, based on its knowledge of both $I$ and the current estimate of $I$ at the receiver. If one wishes to transmit at rates close to capacity, the scheme must be used with a sufficiently large number of interaction rounds, $N$, over the AWGNC. Now, since the constellation size, $M$, is exponentially increasing in $N$ ($M=2^{NR}$), this might lead to an excessive constellation size and extremely small error correction terms computed at the encoder, as will be described later in more detail. As a result, the SK scheme completely breaks down when either the transmitter or receiver are limited to use low precision arithmetic, such as 16 bit floating point numbers (Float16). The numerical issues of the SK scheme were noted by various authors, e.g., \cite{kim2018deepcode}. The case of noisy feedback was also discussed by various authors \cite{kim2007gaussian,noisy_peak_en_const,chance2011concatenated,noisy_fb_rel,ben2017interactive,kim2018deepcode} and will not be considered in this paper.

In this work, we propose a new modified SK communication scheme, that breaks the standard SK transmission protocol into stages, where each stage comprises several SK interaction rounds. In the first stage, based on the available information from the associated SK interaction rounds, the transmitter and receiver agree on a sub-interval that with high probability contains the transmitted PAM symbol representing the message. Then, they both zoom into this decoded sub-interval, and apply additional SK interaction rounds, that eventually enable both parties to further zoom into a finer resolution sub-interval, that (with high probability) contains the transmitted PAM symbol. This process repeats until the message has been completely decoded (the final sub-interval is the decoded symbol). We call our new method a zoom-in SK (ZSK) scheme.
We show that our scheme can practically achieve the same performance (error probability) as standard SK, using low precision arithmetic.

The paper is organized as follows. In section \ref{sec:Defs} we define the setup and introduce notations. We then briefly review the SK scheme, and explain its numerical issues. In section \ref{sec:single_zoom} we describe the proposed scheme for the case of a single zoom-in and analyze its error probability. In Section \ref{sec:multiple_zooms} we extend the method to multiple zoom-in stages and describe an algorithm for determining the zoom-in parameters (number of SK interaction rounds associated with each stage and its constellation size as described below).
In Section \ref{sec:sim} we compare our ZSK scheme with standard SK using computer simulations.

\section{Overview of SK scheme and numerical issues}\label{sec:Defs}
\subsection{Preliminaries} \label{sec:preliminary}
\begin{figure}[H]
	\centering
	\includegraphics[width=0.9\columnwidth]{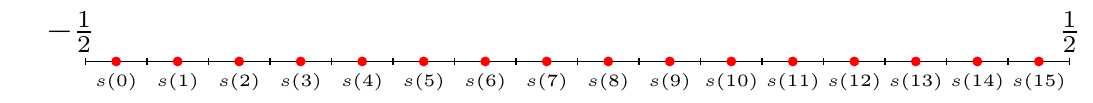}
	\caption{$M$-PAM constellation with $M=16$.}
	\label{fig:pam_const}
\end{figure}

We define the following $M$-PAM constellation, also shown in Figure \ref{fig:pam_const}, that will be used throughout the work,
\bre \label{eq:M_PAM}
\theta(i) = \PAM(i, M) = \frac{i}{M}-\frac{1}{2}+\frac{1}{2M}, \quad i\in [0, M-1].
\ere
Assuming that the input message\footnote{We follow the notation that an upper case letter denotes a random variable (RV), and a lower case letter denotes a particular value that this RV attains.}, $I$, is uniformly distributed, it is straightforward to obtain the average power, $A$, of this constellation as
\begin{equation}
A^2 = \Exp[\Theta(I)^2] = \frac{M^2-1}{12M^2}.
\label{eq:A}
\end{equation}

Suppose that we transmit $\Theta(I)$ over an additive noise channel,
$$
Y = \Theta(I) + Z
$$
where $\Exp[Z]=0$ and $Z$ is distributed symmetrically around zero.
The ML decoding error probability can be easily derived and is given by,
\begin{align}
\label{eq:PAM_error}
P_e^{\PAM} &=
\frac{M-2}{M}\cdot 2\cdot\Pr\left(Z>\frac{1}{2M}\right) + \frac{2}{M}\cdot \Pr\left(Z>\frac{1}{2M}\right)
\\&=
2\left(1-\frac{1}{M}\right)\Pr\left(Z>\frac{1}{2M}\right)
\le 2\Pr\left(Z>\frac{1}{2M}\right)
.
\end{align}
For example, if the noise is Gaussian, i.e., $Z\sim\cN(0,\sigma^2)$, the decoding error probability is given by, 
\bre\label{eq:PAM_error_gaussian}
P_{e, {\rm Gaussian}}^{\PAM}
=
2\left(1-\frac{1}{M}\right)Q\left(\frac{1}{2M\sigma}\right)
\le
2Q\left(\frac{1}{2M\sigma}\right)
\ere
where $Q(x)$, the tail distribution function of the standard normal distribution, is given by
$$
Q(x) \defined \frac{1}{\sqrt{2\pi}} \int_{x}^{\infty} e^{-u^2/2} du
$$
Note that the inequalities in \eqref{eq:PAM_error} and \eqref{eq:PAM_error_gaussian} are actually very tight for non-trivial cases (where $M$ is very small). Hence the upper bounds in these equations are also excellent approximations to the respective error probabilities.
 
In the sequel we use some properties of the linear minimum mean square error (LMMSE) estimator. We highlight some of it's main properties as a reminder. Assume that we are given an input sample, $Y = \frac{\sqrt{P}}{\sigma_x}X+Z$, where $X$ and $Z$ are statistically independent RVs with $\Exp[X]=\Exp[Z]=0$, $\Var[X]=\sigma_x^2$ and $\Var[Z]=\sigma_z^2$. Denote the signal to noise ratio by $\SNR=P / \sigma_z^2$. Given $Y$, we wish to estimate $X$ using a linear estimator, $\widehat{X}^L(Y)$, that minimizes $\Exp[(\widehat{X}^L(Y) - X)^2]$. The estimator is given by \cite{RVs},
\bre \label{eq:LMMSE_estimator}
\widehat{X}^L(Y)= \frac{\Cov(X,Y)}{\Var(Y)}Y = \frac{\frac{\sqrt{P}}{\sigma_x}\sigma_x^2}{P + \sigma_z^2}Y=\frac{\sigma_x}{\sigma_z}\frac{\sqrt{\SNR}}{1+\SNR}Y.
\ere
Denote the estimation error by $E=\widehat{X}^L(Y)-X$. Its variance is given by,
\bre\label{error_var_eq}
\Var(E) = \Var(X)-\frac{\Cov(X,Y)^2}{\Var(Y)} = \frac{\sigma_x^2}{1+\SNR}.
\ere
Moreover, the estimation error is orthogonal to any linear function of the measurements $Y$, i.e., $\E[E\cdot Y] = 0$. We note that if $X$ and $Z$ are both Gaussians, the LMMSE estimator coincides with the general minimum mean square error estimator (MMSE).

\subsection{SK scheme} \label{sec:SK_scheme}
We briefly describe the SK scheme. A detailed explanation can be found in \cite{SK_1,SK_2,SK_3}.

The goal is to reliably transmit a message, $i$, over an AWGNC with feedback, as described in Section \ref{sec:Introduction}, using $N$ interaction rounds (iterations). There are $M=2^{NR}$ possible messages, $i\in[0,M-1]$, where $R$ is the communication rate. We assume an average power constraint $P$ at the channel input.
The modulated PAM symbol, prior to power scaling, is $\theta = \theta(i)$, defined in \eqref{eq:M_PAM}. In the first iteration, the transmitter simply transmits the symbol, normalized to satisfy the input power constraint, i.e., $x_0=\frac{\sqrt{P}}{A}\theta$ and the receiver estimates the transmitted symbol using $\widehat{\theta}_0=\frac{y_0}{\sqrt{P}/A}$, where $y_0$ is the channel output corresponding to $x_0$. In each of the following iterations, the transmitter calculates the receiver's estimation error,
\bre
\epsilon_n=\widehat{\theta}_n-\theta
\label{eq:epsilon_n}
\ere
and transmits it back to the receiver (normalized to satisfy the input power constraint)  $x_{n+1}=\frac{\sqrt{P}}{\sigma_n}\epsilon_n$, where $\sigma_n^2 = \Var(\cE_n)$. The receiver obtains
\bre
y_{n+1} = x_{n+1} + z_{n+1} = \frac{\sqrt{P}}{\sigma_n} \epsilon_n + z_{n+1}
\label{eq:chan_np1}
\ere
where $z_{n+1}$ is the channel noise at the $n+1$'th iteration, and calculates the MMSE estimator of $\epsilon_n$, denoted by $\widehat{\epsilon}_n$ (which, in the Gaussian case, coincides with the LMMSE) using,
\bre
\widehat{\epsilon}_n = \beta_n\cdot y_{n+1}
\label{eq:hat_epsilon_n}
\ere
where $\beta_n$ is the LMMSE estimator coefficient given by \eqref{eq:LMMSE_estimator},
\bre
\beta_n = \frac{\sigma_n}{\sigma_z} \cdot \frac{\sqrt{\SNR}}{1+\SNR}
\label{eq:beta_n}
\ere
It then updates its current estimate using,
\begin{equation}
\widehat{\theta}_{n+1}=\widehat{\theta}_{n}-\widehat{\epsilon}_n.
\end{equation}
Hence, at each round we have, 
\bre
\label{eq:EpsUpdate}
\epsilon_{n+1} = \widehat{\theta}_{n+1}-\theta = \widehat{\theta}_{n}-\widehat{\epsilon}_n-\theta = \epsilon_n-\widehat{\epsilon}_n.
\ere
Thus, the error variance can be recursively calculated using \eqref{error_var_eq},
\bre \label{error_var}
\sigma_{n+1}^2=\frac{\sigma_{n}^2}{1+\SNR}=\frac{\sigma_0^2}{(1+\SNR)^{n+1}}
\ere
where $\sigma_0^2 = A^2 / \SNR$ is the error variance at the first iteration. After $N$ iterations, the symbol is decoded at the receiver using an ML PAM decoder. The decoding is successful if $\left|\cE_{N-1}\right| < \frac{1}{2M}$, and in fact by \eqref{eq:PAM_error_gaussian}, the error is upper bounded (and also well approximated) by,
\begin{align}
P_e^{\SK}
\le
2\Pr\left( \cE_{N-1} \ge \frac{1}{2M} \right)
=
2Q\left(\frac{1}{2M\sigma_{N-1}}\right)
\label{eq:SK}
\end{align}
where the noise variance at the last iteration is given by,
\begin{align}\label{eq:sk_final_snr}
\sigma_{N-1}^{2}=\frac{\sigma_0^2}{(1+\SNR)^{N-1}} = \frac{A^2}{\SNR(1+\SNR)^{N-1}} = \frac{M^2-1}{12M^2\cdot \SNR(1+\SNR)^{N-1}}.
\end{align}
(the last equality is due to \eqref{eq:A}). Thus we have,
\bre
P_e^{\SK} \le 2Q\left(\frac{1}{2M}\sqrt{\frac{12M^2\cdot \SNR(1+\SNR)^{N-1}}{M^2-1}}\right) \le 2Q\left(\sqrt{\frac{3\SNR}{1+\SNR}\frac{(1+\SNR)^N}{M^2}}\right).
\ere
Plugging in $C=\half\log_2(1+\SNR)$, and $M=2^{NR}$, we have,
\bre \label{eq:SK_error}
P_e^{\SK} \le 2Q\left(\sqrt{\frac{3\SNR}{1+\SNR}\cdot2^{2N(C-R)}}\right).
\ere
which is the well known SK error probability.
The SK scheme is summarized in Algorithm \ref{alg:SK}.
\begin{algorithm}[H]
	\caption{SK}
	\label{alg:SK}
	\begin{algorithmic}
		\Procedure {SK}{$i$: message}
		\State \textbf{Initialize:}
		\State \hspace*{\algorithmicindent} Transmitter: $\theta=\PAM(i,M)$, $x_{0}=\frac{\sqrt{P}}{A}\theta$
		\State \hspace*{\algorithmicindent} Receiver: $\widehat{\theta}_0=\frac{y_0}{\sqrt{P}/A}$
		\For{$n=0,\ldots,N-2$}
		\State  Transmitter: $\epsilon_{n}=\widehat{\theta}_n-\theta$
		\State  $x_{n+1}=\frac{\sqrt{P}}{\sigma_n}\epsilon_n$
		\State   Receiver: $\widehat{\epsilon}_n=\beta_n\cdot y_{n+1} = \beta_n\cdot \left( x_{n+1} + z_{n+1} \right)$
		\State   $\widehat{\theta}_{n+1}=\widehat{\theta}_n-\widehat{\epsilon}_n$
		\EndFor
		\Return$\widehat{i}=\argmin_l\{||\widehat{\theta}_{N-1}-\PAM(l, M)||^2 \}$
		\EndProcedure
	\end{algorithmic}
\end{algorithm}

\subsection{Numerical issues} \label{sec:numeric}
Many of toady's practical receivers use Float16 \cite{float16_standard} as their main variable for digital signal processing (DSP) calculations. Unfortunately, the use of this low precision variable with the SK iterative feedback decoding scheme is impossible even for a moderate number of iterations or rate. There are mainly two issues:
\begin{enumerate}
	\item The error variance $\sigma_n$ decreases exponentially fast to zero as can be seen in \eqref{error_var}. Thus, $\sigma_n$ vanishes quickly when using low precision representation such as Float16. This phenomenon affects almost all calculations in SK, as $\beta_n$ and the transmission normalization factor $\frac{\sigma_n}{\sqrt{P}}$ are proportional to $\sigma_n$.
	\item Increasing the number of iterations, $N$, will decrease the capacity gap and / or decrease the error rate, as can be seen in \eqref{eq:SK_error}. However, increasing $N$ will also increase the constellation size exponentially fast, as $M=2^{NR}$. As a result, low precision arithmetic such as Float16 may be insufficient to represent the distance of $1/M$ between two adjacent symbols. That is, using Float16 causes aliasing and an error floor.
\end{enumerate}
As a result, the iterative SK scheme fails under low precision arithmetic (Float16) even for moderate values of $N$ and $R$, as can be seen in Figure \ref{fig:float16_SK_no_zoom}. Even when using Float32 the scheme fails when we try to use a large number of iterations or a high rate, as can be seen in Figure \ref{fig:float32_SK_no_zoom}. Note that simple solutions, such as storing the logarithms of the variables in the SK scheme and operating on them are not sufficient for solving the issues indicated above. The numerical issues with the SK scheme have been noted before, e.g. \cite{kim2018deepcode}.

\section{Single zoom-in scheme}\label{sec:single_zoom}
\subsection{The new algorithm} \label{sec:single_zoomin_alg}
To overcome the numerical issues described above, we propose a new modified SK communication scheme, termed zoom-in SK (ZSK). We start with a single zoom-in scheme, that breaks the standard SK transmission protocol into two stages. In the next section we generalize the method to an arbitrary number of zoom-in stages.

Let $M$ be written as $M = M_0 \cdot M_1$.
The idea of the proposed scheme is to break the decoding into two stages. In the first stage, the transmitter and receiver start by applying $k+1$ standard SK interaction rounds (the first round is the initialization one, see Algorithm \ref{alg:SK}). Denote by $\widehat{\theta}_k$ the estimate of the transmitted PAM message $\theta$ after the $k+1$ interaction rounds. Instead of decoding $\theta$ based on $\widehat{\theta}_k$, the receiver just determines an interval $[a,b] \in [-1/2,1/2]$ of size $1/M_0$ (i.e., $b-a=1/M_0$) that contains $\theta$ with high probability. The transmitter, that knows everything about the receiver due to the feedback, makes the same decision.
In the second stage, the transmitter and receiver zoom into the interval $[a,b]$ synchronously (as described below), and apply $N-k-1$ additional interaction rounds, so that in the end of this stage the receiver can decode $\theta$ with high probability.

\begin{figure}[H]
\centering
\includegraphics[width=0.9\columnwidth]{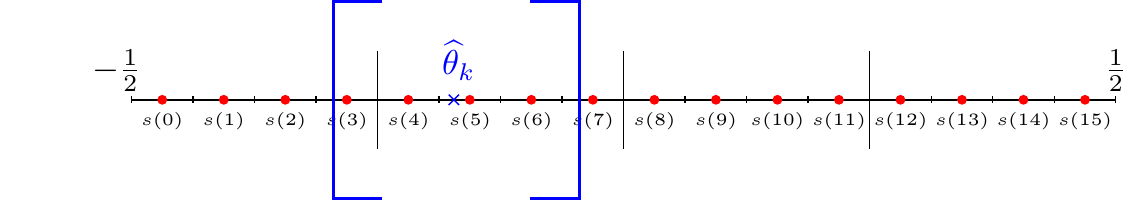} 
\caption{The ZSK scheme with a single zoom-in for $M=16$, $M_0=4$ and $M_1=4$. The original $M$ constellation points are marked by red dots. The estimate $\widehat{\theta}_k$ at the receiver after $k+1$ initial SK interaction rounds is denoted by blue `$\times$', and the initial interval $S'_1$ is denoted by a pair of blue square brackets.}
\label{fig:zoom_in_segment}
\end{figure}

\begin{figure}[H]
\centering
\includegraphics[width=0.9\columnwidth]{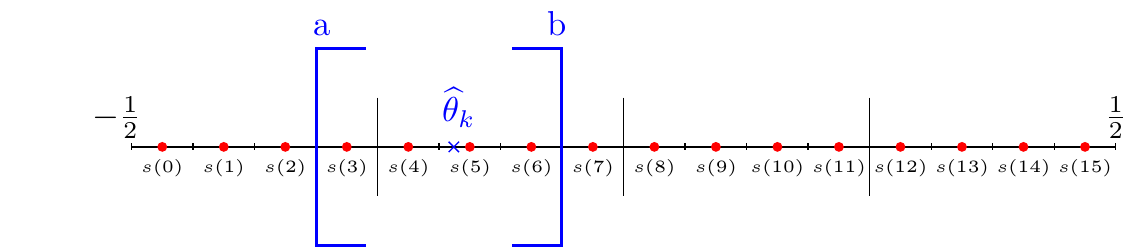} 
\caption{Aligned zoom segment, $S_1=[a,b]$. Here $i'_0=3$.}
\label{fig:aligned_zoom_a_b}
\end{figure}

The interval $[a,b]$ is determined as follows.
First, the receiver constructs the interval
$S'_1 = \left[\widehat{\theta}_k-\frac{1}{2M_0}, \widehat{\theta}_k+\frac{1}{2M_0}\right]$
of size $\frac{1}{M_0}$ around the current estimate, $\widehat{\theta}_k$. For example, we plotted this initial interval for the case where $M=16$, $M_0=4$ and $M_1=4$ in Figure \ref{fig:zoom_in_segment}.
Then, the receiver aligns the interval $S'_1$ with the original PAM constellation of size $M$ (see Fig. \ref{fig:pam_const}) by first computing the number of symbols that are on the left of $S'_1$, denoted by $i'_0$, i.e.,
\begin{equation}
i'_0={\rm round}\left(\left(\widehat{\theta}_k-\frac{1}{2M_0}+\frac{1}{2}\right)\cdot M \right).
\label{eq:ip0_round}
\end{equation}
In the example shown in Figure \ref{fig:zoom_in_segment}, we have, $i'_0=3$.
Then the receiver applies $[0,M-M_1]$-clipping on $i'_0$:
\begin{equation}
i_0=\min\left(M-M_1,\max(i'_0,0)\right).
\label{eq:i0_clip}
\end{equation}
The receiver stores $i_0$ in its memory.
We note here that the only variables that need to be kept with a high enough resolution at the transmitter and receiver are the transmitted symbol $i$ and the decoded symbol (naturally if we want to transmit and decode a $K=N\cdot R$-bit word we need a $K$-bit variable in memory). These are stored as integers. Instead of storing the constellation size $M$ we store its logarithm $NR$ as an integer.
We then align the interval $S'_1$ by constructing the interval $S_1=[a, b]$, as can be seen in Figure \ref{fig:aligned_zoom_a_b},  where
\begin{align}
a&=i_0\cdot\frac{1}{M}-\frac{1}{2}, \label{eq:a_from_i1}\\
b&=a+\frac{1}{M_0}.
\end{align}
Finally, both the transmitter and receiver zoom into the interval $S_1$ by updating the current estimate $\widehat{\theta}_k$ to $\widehat{\theta}_k^{(1)}$ using a simple linear transformation,
\begin{equation}
\widehat{\theta}_k^{(1)} = \frac{\widehat{\theta}_k-a}{b-a}-\frac{1}{2} = M_0\cdot(\widehat{\theta}_k-a) -\frac{1}{2}.
\end{equation}
Hence, after zooming in, $\widehat{\theta}_k=a$ ($\widehat{\theta}_k=b$, respectively) is transformed to $\widehat{\theta}_k^{(1)}=-1/2$ ($\widehat{\theta}_k^{(1)}=1/2$), so that the interval $S_1=[a,b]$ has been transformed to the interval $[-1/2,1/2]$.
Define
$$
\theta^{(1)} \defined M_0 (\theta-a) - \half
$$
In the second stage of our ZSK scheme, we replace the decoding of $\theta \in S_1 = [a,b]$ by the decoding of $\theta^{(1)} \in [-1/2,1/2]$. This is done by applying $N-k-1$ additional SK interaction rounds, starting with the current zoom-in estimate $\widehat{\theta}_k^{(1)}$ to $\theta^{(1)}$. 
Recalling that at the $k$'th iteration, $\widehat{\theta}_k=\theta+\epsilon_k$ (see \eqref{eq:epsilon_n}), we have,
\begin{equation} \label{eq:t_k_zoom}
\widehat{\theta}_k^{(1)} = M_0\cdot\left(\theta+\epsilon_k-a\right)-\frac{1}{2} = 
M_0 (\theta - a)-\frac{1}{2}+M_0\epsilon_k = \theta^{(1)} + M_0\epsilon_k.
\end{equation}
\begin{lemma}
Defining $i_1 \defined i-i_0$, we have
\bre
\theta^{(1)} = \PAM(i_1, M_1)=\frac{i_1}{M_1}+\frac{1}{2M_1}-\half.
\label{eq:PAM_i1_M1}
\ere
\label{lem:theta1_i2}
\end{lemma}
\beginproof
\begin{align}
\theta^{(1)}=&M_0(\theta-a)-\frac{1}{2}=M_0(\theta -\frac{i_0}{M}+\frac{1}{2})-\half\\ 
=&
M_0(\frac{i}{M}-\half+\frac{1   }{2M}-\frac{i_0}{M}+\half)-\half=\\  
=&
M_0(\frac{i-i_0}{M}+\frac{1}{2M})-\half = \frac{i-i_0}{M_1}+\frac{1}{2M_1}-\half \\
=& 
\frac{i_1}{M_1}+\frac{1}{2M_1}-\half.
\end{align}
\finproof
Now, after the zoom-in operation, when $i_0$ has already been decoded, it remains to decode $i_1$ in order to conclude the decoding of $i=i_0+i_1$. By Lemma \ref{lem:theta1_i2}, the decoding of $i_1$ is equivalent to the decoding of the PAM symbol $\theta^{(1)}$ corresponding to $i_1$, for a constellation size of $M_1$. We implement the decoding of $i_1$ by using $N-k-1$ SK interaction rounds in the second stage of the ZSK scheme. Our initial estimate to $\theta^{(1)}$ in the second stage of ZSK is $\widehat{\theta}^{(1)}_k$. Our estimate to $\theta^{(1)}$ at the $n$'th interaction round of ZSK, for $n=k,k+1,\ldots,N-1$, is $\widehat{\theta}^{(1)}_n$. We also denote the respective estimation error by $\epsilon_{n}^{(1)}=\widehat{\theta}_n^{(1)}-\theta^{(1)}$. Motivated by the increased error variance in the second stage (as seen in \eqref{eq:t_k_zoom}), we suggest the following updated parameters for the second stage of ZSK, 
\begin{align}
\label{eq:SecondStageParams}
(\sigma_{k}^{(1)})^2 &= M_0^2 \sigma_{k}^2\\
(\sigma_{n}^{(1)})^2 &= \frac{(\sigma_{n-1}^{(1)})^2}{1+\SNR}
\quad
n=k+1,\ldots,N-2\\
\beta_{n}^{(1)} &= \frac{\sqrt{\SNR}}{1+\SNR}\frac{\sigma_{n}^{(1)}}{\sigma_z}
\quad
n=k,\ldots,N-2
\: .
\end{align}

A summary of the ZSK scheme is provided in Algorithm \ref{alg:ZSK}.

\begin{algorithm}[H]
	\caption{ZSK}
	\label{alg:ZSK}
	\begin{algorithmic}
		\Procedure {ZSK}{$i$: message}
		\State \textbf{Initialize:}
		\State \hspace*{\algorithmicindent} Transmitter: $\theta=\PAM(i,M)$, $x_{0}=\frac{\sqrt{P}}{A}\theta$
		\State \hspace*{\algorithmicindent} Receiver: $\widehat{\theta}_0=\frac{y_0}{\sqrt{P}/A}$
		\For{$n=0,\ldots,k-1$}
		\State  Transmitter: $\epsilon_{n}=\widehat{\theta}_n-\theta$
		\State  $x_{n+1}=\frac{\sqrt{P}}{\sigma_n}\epsilon_n$
		\State   Receiver: $\widehat{\epsilon}_n=\beta_n\cdot y_{n+1} = \beta_n\cdot \left( x_{n+1} + z_{n+1} \right)$
		\State   $\widehat{\theta}_{n+1}=\widehat{\theta}_n-\widehat{\epsilon}_n$
		\EndFor
		\State \textbf{Zoom} transmitter and receiver:
		\State $i'_0={\rm round}\left(\left(\widehat{\theta}_k-\frac{1}{2M_0}+\frac{1}{2}\right)\cdot M \right)$
		\State $i_0=\min\left(M-M_1,\max(i'_0,0)\right)$
		\State $a=i_0\cdot\frac{1}{M}-\frac{1}{2}$
		\State $\widehat{\theta}_k^{(1)} = M_0\cdot(\widehat{\theta}_k-a) -\frac{1}{2}$
		\State  Transmitter: $i_1=i-i_0$, $\theta^{(1)}=\PAM(i_1,M_1)$
		\For{$n=k,\ldots,N-2$}
		\State  Transmitter: $\epsilon_{n}^{(1)}=\widehat{\theta}_n^{(1)}-\theta^{(1)}$
		\State  $x_{n+1}=\frac{\sqrt{P}}{\sigma_n^{(1)}}\epsilon_n^{(1)}$
		\State   Receiver: $\widehat{\epsilon}_n^{(1)}=\beta_n^{(1)}\cdot y_{n+1} = \beta_n^{(1)}\cdot \left( x_{n+1} + z_{n+1} \right)$
		\State   $\widehat{\theta}_{n+1}^{(1)}=\widehat{\theta}_n^{(1)}-\widehat{\epsilon}_n^{(1)}$
		\EndFor
		
		\State$i_1=\argmin_l\{||\widehat{\theta}_{N-1}^{(1)}-\PAM(l, M_1)||^2 \}$
		\Return$\widehat{i}=i_0+i_1$
		\EndProcedure
	\end{algorithmic}
\end{algorithm}
As a result of the zoom-in operation, the error variance increases, but at the same time the constellation size decreases such that the numerical robustness is improved while having negligible effect on the error probability (as will be seen in the next section). The numerical robustness improves, mainly because the ZSK scheme allows us to keep the error variance (and thus all the other variables which are linearly related to it) relatively high, such that they can be represented using low precision. Assuming that $M=M_{0}\cdot M_{1}$, while in SK we need $\sigma_{N-1}\ll 1/M$ at the final iteration, for ZSK we require $\sigma_{k}\ll1/M_0$, thus saving approximately $\log_{2}(M/M_0)=\log_{2}(M_1)$ bits in representation. If we assume for simplicity that $M_0=M_1=\sqrt{M}$, then in ZSK we reduced by half the number of bits needed in order to represent the error variance. Even though we might have a small numerical error in decoding $i_0$, the decoded value will be the same in the receiver and transmitter, so that they both stay synchronized, and thus this small error will not pose a problem.

\subsection{Error probability analysis} \label{sec:error_analysis}
As was discussed in Section \ref{sec:SK_scheme}, the SK scheme decodes successfully if $\left|\cE_{N-1}\right| < \frac{1}{2M}$.
By the discussion in Section \ref{sec:single_zoomin_alg}, the first stage decoding of the ZSK scheme is successful if
\begin{equation}
\label{eq:noErrorFirst}
\Theta \in \left[\widehat{\Theta}_k-\frac{1}{2M_0}, \widehat{\Theta}_k+\frac{1}{2M_0}\right]
\end{equation}
This event is equivalent to $\left| \cE_k \right| < 1/[2M_0]$.
By Lemma \ref{lem:theta1_i2}, the second stage decoding of the ZSK scheme is successful if
\begin{equation}
\label{eq:noErrorSecond}
\Theta^{(1)} \in 
\left[\widehat{\Theta}_{N-1}^{(1)} - \frac{1}{2M_1}, \widehat{\Theta}_{N-1}^{(1)}+\frac{1}{2M_1}\right]
\end{equation}
This event is equivalent to $\left| \cE_{N-1}^{(1)} \right| < 1/[2M_1]$.

The error probability of the ZSK scheme can be analyzed using truncated RVs. However, a simpler analysis is provided below using a coupling argument similar to the one used in \cite{ben2017interactive} in the context of noisy feedback.  

\begin{theorem} \label{th:err_prb}
The ZSK error probability is upper bounded by the sum of the zoom error probability and the regular SK error probability, 
\bre
P_{e}^{\ZSK} \le
2Q\left(\frac{1}{2M_0\sigma_{k}}\right) + 2Q\left(\frac{1}{2M\sigma_{N-1}}\right)
\label{eq:th_err_prb}
\ere
\end{theorem}
As an immediate corollary we have:
\begin{corollary} \label{corr:err_prb}
Suppose that $M_0$ and $k$ are chosen such that
$Q\left(\frac{1}{2M_0\sigma_{k}}\right) < \epsilon Q\left(\frac{1}{2M\sigma_{N-1}}\right)$
for some (small) $\epsilon>0$. Then,
\bre
P_{e}^{\ZSK} \le 2(1+\epsilon) Q\left(\frac{1}{2M\sigma_{N-1}}\right)
\ere
\end{corollary}
As will be seen in Section \ref{sec:choosing_params}, we set the parameters of the ZSK scheme, which in the single zoom case are $M_0$ and $k$, such that the required condition in Corollary \ref{corr:err_prb} is satisfied for small $\epsilon>0$.
It can be seen that the bound on $P_{e}^{\ZSK}$ in Corollary \ref{corr:err_prb} is essentially (up to $1+\epsilon$) the same as the bound in \eqref{eq:SK} on the SK error probability, $P_e^{\SK}$, with the same total number of iterations, $N$. Furthermore, as was noted above, the bound \eqref{eq:SK} is an excellent approximation to $P_e^{\SK}$. Hence, under a proper design of the ZSK scheme, its error probability is essentially the same as that of plain SK.

{\noindent \textbf{Proof of Theorem \ref{th:err_prb}: }  }
Consider two systems that are fed with the exact same message and experience the exact same channel noises. The first one applies the proposed ZSK algorithm, while the second one applies plain SK. The parameters and signals of the SK system are denoted by $\sigma_n$, $\beta_n$, $\epsilon_n$, $\widehat{\epsilon}_n$ and $y_{n+1}$. The same parameters and signals are used by the ZSK system for $n=0,\ldots,k-1$ (before the zoom in). The parameters and signals of the ZSK scheme after the zoom in are denoted by $\sigma_n^{(1)}$, $\beta_n^{(1)}$, $\epsilon_n^{(1)}$, $\widehat{\epsilon}_n^{(1)}$ and $y^{(1)}_{n+1}$ for $n=k,\ldots, N-2$.
We claim that if $|\epsilon_k| < 1 / [2 M_0]$ and $|\epsilon_{N-1}| < 1 / [2 M]$ then both systems will decode the transmitted message successfully.
If this claim indeed holds then
\begin{equation}
P_e^{\ZSK} \le 
\Pr \left\{  \left|\cE_k\right| \ge \frac{1}{2M_0} \bigcup \left|\cE_{N-1}\right| \ge \frac{1}{2M} \right\}
\end{equation}
which immediately proves \eqref{eq:th_err_prb} by the union bound and the analysis of the SK scheme in Section \ref{sec:SK_scheme} (see \eqref{eq:SK} for the second term on the right hand side of \eqref{eq:th_err_prb}, and the same argument can also be used to obtain the first term on the right hand side of \eqref{eq:th_err_prb}).

Now, the above claim obviously holds for the SK system since $|\epsilon_{N-1}| < 1 / [2 M]$.
It remains to prove the claim for the ZSK system.
The first stage of ZSK decoding is successful since by assumption, $|\epsilon_k| < 1 / [2 M_0]$. We show that the second stage of ZSK decoding is also successful by showing that the second assumption of the claim, $|\epsilon_{N-1}| < 1 / [2 M]$, is equivalent to $\left| \epsilon_{N-1}^{(1)} \right| < 1/[2M_1]$. 
For that, it is sufficient to show that given $\left| \epsilon_k \right| < 1/[2M_0]$, so that the first stage decoding of the ZSK scheme was successful, we have 
\begin{equation}
\epsilon_n^{(1)} = M_{0}\epsilon_{n}
\label{eq:epsilon_M0}
\end{equation}
for $n=k,\ldots,N-1$ (i.e., in the second stage of the ZSK scheme the estimation errors are $M_0$ times larger than the corresponding error in the SK scheme). For $n=k$ \eqref{eq:epsilon_M0} holds by \eqref{eq:t_k_zoom}. We proceed by induction: Suppose that \eqref{eq:epsilon_M0} holds for $n=r$. Then, by \eqref{eq:EpsUpdate}, \eqref{eq:SecondStageParams}, Algorithm \ref{alg:SK}, Algorithm \ref{alg:ZSK} and the induction assumption, 
\begin{align}
\epsilon_{r+1}
&= \epsilon_{r} - \beta_{r}y_{r+1}\\
&= \epsilon_{r} - \frac{\sigma_r}{\sigma_z} \cdot \frac{\sqrt{\SNR}}{1+\SNR}\left(\frac{\sqrt{P}}{\sigma_r}\epsilon_{r}+z_{r+1}\right)\\
\epsilon_{r+1}^{(1)} &= \epsilon_{r}^{(1)} - \beta_{r}^{(1)}y_{r+1}^{(1)}\\
&= M_{0}\epsilon_{r} - \frac{M_{0}\sigma_r}{\sigma_z} \cdot \frac{\sqrt{\SNR}}{1+\SNR}\left(\frac{\sqrt{P}}{M_{0}\sigma_r}M_{0}\epsilon_{r}+z_{r+1}\right)\\
&= M_{0}\epsilon_{r+1}
\end{align}
This concludes the induction, the proof of the claim, and the proof of the theorem.

\finproof

\section{Multiple zooms} \label{sec:multiple_zooms}
In the previous section we have described how the zoom scheme works for the case of a single zoom. It is straight forward to generalize it into a scheme with multiple zooms where the transmitter and receiver zoom synchronously every few iterations. This way we can implement an SK scheme with an arbitrarily large number of iterations and still use low precision arithmetic. Consider a multiple zoom SK scheme with $r$ zoom-ins and $r+1$ stages (such that for $r=1$ it reduces to the single zoom-in case with 2 stages discussed earlier). Suppose that $M$ can be written as
\bre
M = \prod_{j=0}^{r} M_j
\ere
and that the $j$'th zoom-in operation, $j=0,\ldots,r-1$, is performed after $k_{j} + 1$ interaction rounds. The last zoom-in is performed after $k_{r-1}+1$ interaction rounds. Immediately after the last zoom-in we carry out the last $N-k_{r-1}-1$ interaction rounds for a total of $N$ interaction rounds. We also define $k_r \defined N-1$.
The estimation error random variables at the $j$'th stage ($j=0,1,\ldots,r$) of ZSK are denoted by $\cE_n^{(j)}$, where $n$ is the interaction round index, $n=k_{j-1},k_{j-1}+1,\ldots,k_{j}$ and $k_{-1}\equiv 0$. Before the $j$'th zoom-in the estimation error is $\cE_{k_j}^{(j)}$, and after the zoom-in it is $\cE_{k_j}^{(j+1)}$. Similarly to \eqref{eq:SecondStageParams}, we suggest the following updated parameters for the $j$'th stage,
\begin{align}
\label{eq:SecondStageParams1}
(\sigma_{k_{j-1}}^{(j)})^2 &= \left[\prod_{l=0}^{j-1}M_l^2\right] \sigma_{k_{j-1}}^2\\
(\sigma_{n}^{(j)})^2 &= \frac{(\sigma_{n-1}^{(j)})^2}{1+\SNR}
\quad
n=k_{j-1}+1,\ldots,k_j-1\\
\beta_{n}^{(j)} &= \frac{\sqrt{\SNR}}{1+\SNR}\frac{\sigma_{n}^{(j)}}{\sigma_z}
\quad
n=k_{j-1},\ldots,k_j-1
\: .
\end{align}
where $\sigma_k^2$ is the SK error variance at the $k$'th iteration given by \eqref{error_var}. Similarly to \eqref{eq:noErrorFirst}-\eqref{eq:noErrorSecond} a zoom error event at the $j$'th stage is equivalent to the event $\left|\cE_{k_j}^{(j)}\right|>1/\left[2M_j\right]$. As an example consider the single zoom-in case where $r=1$. In this case there is a single zoom-in after $k_0+1$ interaction rounds (in the previous section, where we considered the single zoom-in case, $k_0$ was denoted by $k$). During the $0$'th stage of ZSK, the estimation errors are $\cE_{n}^{(0)}$, for $n=0,1,\ldots,k_0$ and an error at the end of that stage is equivalent to the event $\left|\cE_{k_0}^{(0)}\right| > 1/\left[2M_0\right]$ (in the previous section the superscript $(0)$ was omitted for the $0$'th stage of ZSK). After the zoom-in the estimation errors are $\cE_{n}^{(1)}$, for $n=k_0,k_0+1,\ldots,k_1$, and an error at the end of that stage is equivalent to the event $\left|\cE_{k_1}^{(1)}\right| > 1/\left[2M_1\right]$ where $k_1=N-1$.

As an extension of Theorem \ref{th:err_prb} to the multiple zooms case we have the following.
\begin{theorem} \label{th:err_prb_MZSK}
The multiple ZSK error probability is upper bounded by,
\bre
P_{e}^{\ZSK} \le \sum_{j=0}^{r}
2Q\left(\frac{1}{2 M_{0}\ldots M_{j} \sigma_{k_{j}}}\right)
\label{eq:th_err_prb_MZSK}
\ere
where $\sigma_n^2$ is the standard SK error variance at the $n$'th interaction round (as given in \eqref{error_var}).
\end{theorem}    
As an immediate corollary we have:
\begin{corollary} \label{corr:err_prb_MZSK}
Suppose that $\{M_j\}_{j=0}^r$ and $\{k_j\}_{j=0}^r$ are chosen such that
$$
Q\left(\frac{1}{2\cdot M_{0}\ldots M_{j} \sigma_{k_{j}}}\right) < \epsilon Q\left(\frac{1}{2M\sigma_{N-1}}\right)
$$
for $j=0,\ldots,r-1$. Then
\bre
P_{e}^{\ZSK} \le 2(1+r\epsilon) Q\left(\frac{1}{2M\sigma_{N-1}}\right)
\ere
\end{corollary}
As will be seen in Section \ref{sec:choosing_params}, we set the parameters of the ZSK scheme, $M_j$ and $k_j$, such that the required condition in Corollary \ref{corr:err_prb_MZSK} is satisfied for small $\epsilon>0$.
It can be seen that the bound on $P_{e}^{\ZSK}$ in Corollary \ref{corr:err_prb_MZSK} is essentially (up to $1+r\epsilon$) the same as the bound in \eqref{eq:SK} on the SK error probability, $P_e^{\SK}$, with the same total number of iterations, $N$. Furthermore, as was noted above, the bound \eqref{eq:SK} is an excellent approximation to $P_e^{\SK}$. Hence, under a proper design of the multiple ZSK scheme, its error probability is essentially the same as that of plain SK.

{\noindent \textbf{Proof of Theorem \ref{th:err_prb_MZSK}: }  }
Similarly to the proof of Theorem \ref{th:err_prb}, we compare two systems which are fed with the same message and experience the same noises. The first applies the proposed multiple stage ZSK algorithm while the second applies plain SK. Using the same notation (where the estimation errors in the SK system are denoted by $\epsilon_n$, and in the ZSK system they are marked with an additional superscript indicating the stage), we claim that if for $j=0,1,\ldots,r$ we have,  
\bre
\label{eq:noZoomError}
\left|\epsilon_{k_j}\right|<\left[2\prod_{l=0}^{j} M_l\right]^{-1}
\ere
then both systems will decode the transmitted message successfully. If this claim indeed holds then the ZSK error probability is upper bounded by, 
\begin{equation}
P_e^{\ZSK} \le 
\Pr \left( \bigcup_{j=0}^{r}\left\{\left|\cE_{k_j}\right| \ge \left[2\prod_{l=0}^{j} M_l\right]^{-1}\right\}\right)
\end{equation}
which immediately proves the theorem by the union bound and the analysis of the SK scheme in Section \ref{sec:SK_scheme}.
Now, the claim obviously holds for the SK system since $|\epsilon_{k_r}| < 1 / [2 M]$. Next, we show that the event \eqref{eq:noZoomError} implies a successful zoom at all the zoom steps and a successful decoding at the last iteration. It can be seen that by the exact same arguments as in Theorem \ref{th:err_prb}, given that \eqref{eq:noZoomError} holds for $j=0, \ldots, j_0-1$ then at the $j_0$'th stage we have, 
\bre
\epsilon_n^{(j_0)} = \epsilon_n\prod_{i=0}^{j_0-1} M_{i} \quad n=k_{j_0-1}, \ldots,k_{j_0}
\ere
which implies that the event $\left|\epsilon_{k_{j_0}}\right| \ge \left[2\prod_{l=0}^{j_0} M_l\right]^{-1}$ is equivalent to $\epsilon_{k_{j_0}}^{(j_0)}>\left[2M_{j_0}\right]^{-1}$. But this is exactly the error event at the $j_0$'th stage of the ZSK scheme. Thus, we see that the event \eqref{eq:noZoomError} indeed implies a successful decoding in the ZSK scheme as well. As a result, the claim holds and the theorem follows.

\finproof

\subsection{Choosing zoom parameters} \label{sec:choosing_params}
It remains to show how we determine the zoom constellation sizes,
$\bM \defined (M_0,M_1,\ldots,M_{r})$ and the iteration indices to zoom at, $\bK \defined (k_0,k_1,\ldots,k_{r-1})$, so that the total error probability of ZSK will be essentially the same as the error probability of standard SK, without the numerical issues of standard SK. First, we set a target error probability $P_e^{\target}$ that we want to achieve with SK (e.g., $P_e^{\target}=10^{-6}$). Then, we calculate at what SNR the standard SK scheme will reach that error probability by solving,
\begin{align}
P_e^{\target} &= 2Q\left(\frac{1}{2M\sigma_{N-1}}\right),\\
\sigma^2_{N-1} &= \frac{\sigma^2_0}{(1+\SNR)^{N-1}}.
\end{align}
We denote that SNR by $\SNR_{\target}$. Next, we set the desired error probability at each zoom step, $P_e^{\zoom}=\epsilon P_e^{\target}$ (e.g., with $\epsilon=10^{-3}$), such that the sum of all zoom errors will have a negligible effect on the final error probability. We calculate the error variance at each iteration when the SNR is $\SNR_{\target}$, for the standard SK scheme and store it in the array $\bsigma_{\ZSK}^2[\iter]=\frac{\sigma_0^2}{(1+\SNR_{\target})^{\iter}}$ for $\iter=0,\ldots,N-1$ (this is done off-line so we can store these values (or their logarithms) at any desired accuracy).
Next, we can use the following algorithm to set the iterations at which we need to zoom-in and the corresponding zoom constellation.
\begin{algorithm}[H]
	\caption{Finding zoom parameters}
	\begin{algorithmic}
		\Procedure {Find zoom parameters}{$P_e^{\zoom}$: upper bound on the zoom error probability, $\bsigma_{\ZSK}^2$: error variances of plain SK at $\SNR_{\target}$}
		\For{$i = 1,2,\ldots,N-1$}
		\State $\sigma^2=\bsigma_{\ZSK}^2[i]$
		\For {NumBits $=\log_2 M,\dots,1$}
		\State $M_{z} = 2^{{\rm NumBits}}$
		\State$P_{e} = 2Q(\frac{1}{2M_{z}\sigma})$
		\If{$P_e<P_e^{\zoom}$}
		\State $\bK$.append($i$)
		\State $\bM$.append($M_{z}$)
		\State $\bsigma_{\ZSK}^2=\bsigma_{\ZSK}^2\cdot M_{z}^2$
		\State$M=M/M_{z}$
		\State Break
		\State
		\EndIf
		\EndFor
		\EndFor
		\Return $\bK,\bM$ 
		\EndProcedure
	\end{algorithmic}
\end{algorithm}
At each iteration ($i$), the algorithm tests whether there exists a constellation size, such that zooming in to that constellation at iteration $i$ will result in a zoom error that is smaller than $P_e^{\zoom}$. If such constellation exists, it chooses the maximal $M_{z}$ possible and updates $M$ and the error variances $\bsigma_{\ZSK}^2$.

\section{Simulation results}\label{sec:sim}
\begin{figure}[H]
	\centering
	\includegraphics[scale=0.5]{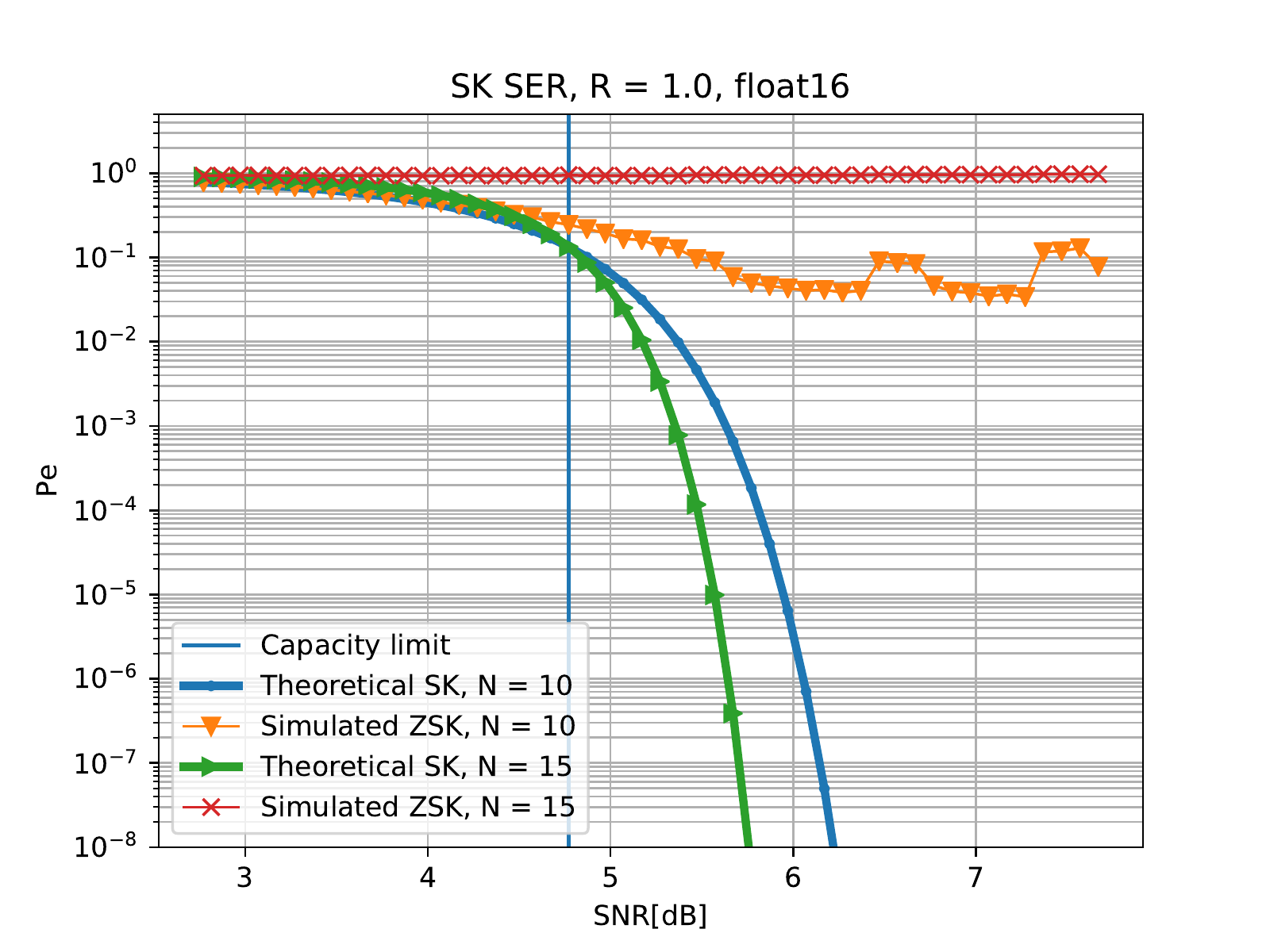}
	\caption{The symbol error rate of the standard SK scheme with Float16.}
	\label{fig:float16_SK_no_zoom}
\end{figure}

\begin{figure}[H]
	\centering
	\includegraphics[scale=0.5]{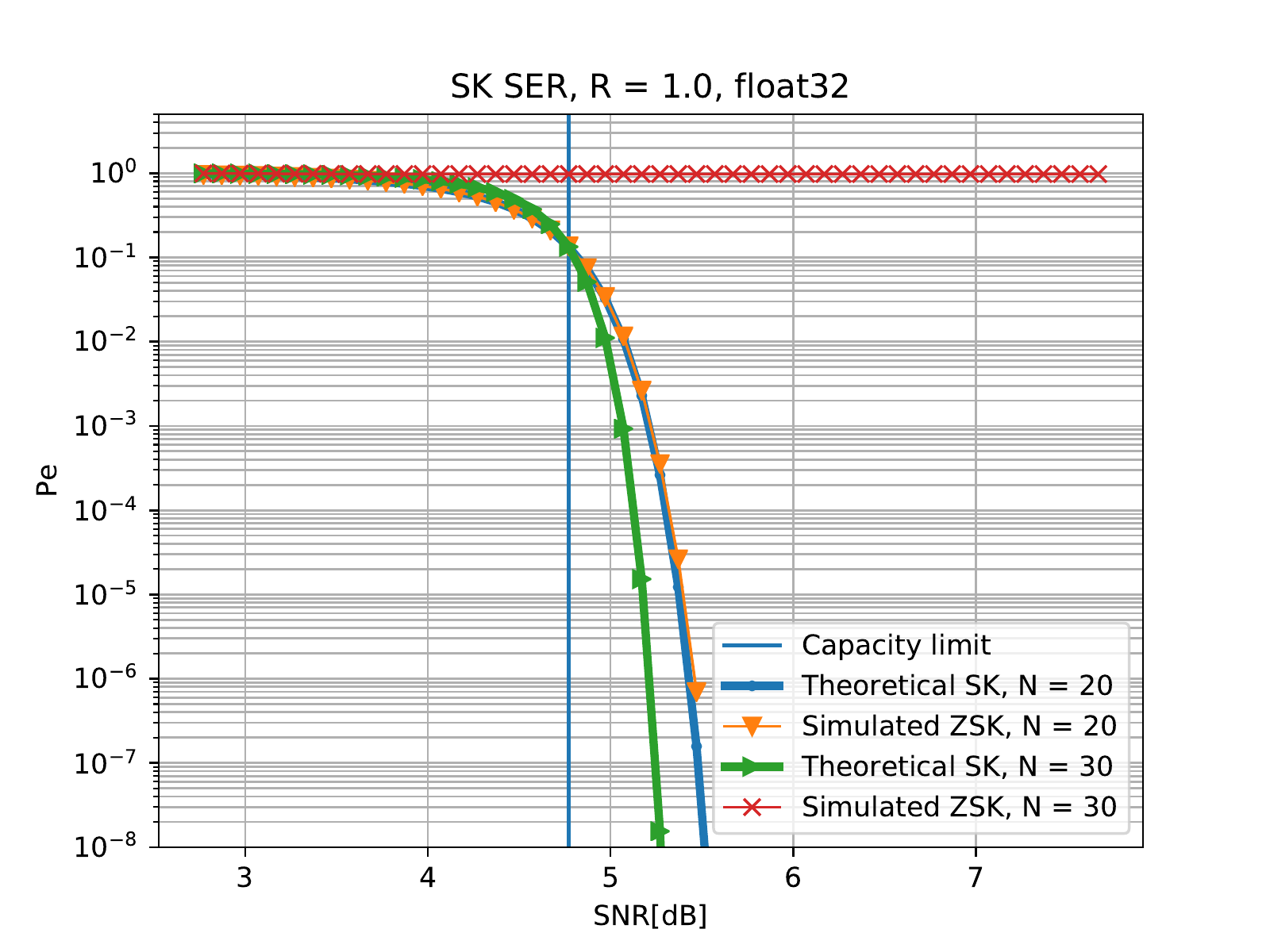}
	\caption{The symbol error rate of the standard SK scheme with Float32.}
	\label{fig:float32_SK_no_zoom}
\end{figure}

\begin{figure}[H]
	\centering
	\includegraphics[scale=0.5]{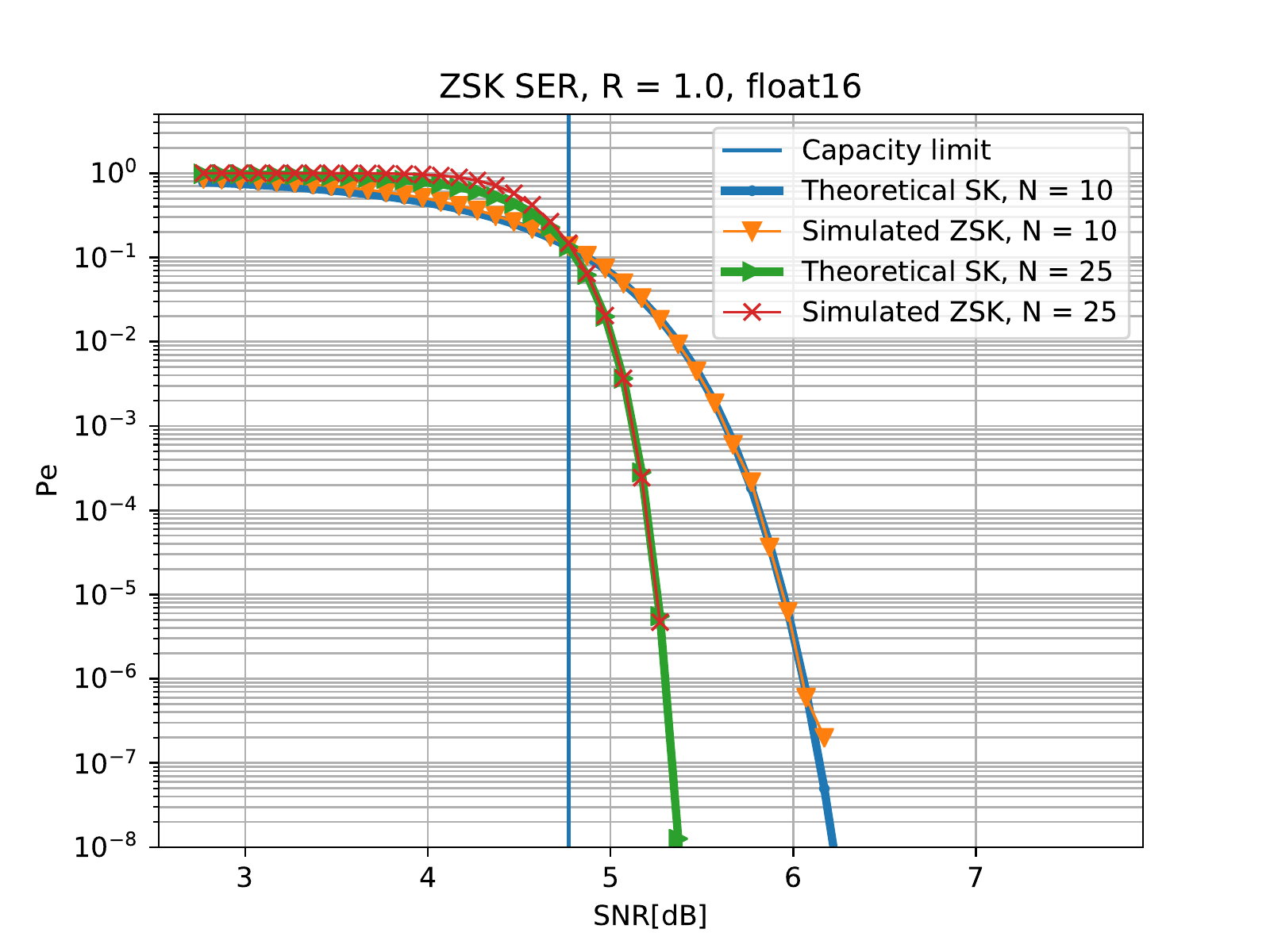}
	\caption{Zoom scheme symbol error rate for $N=10$ with $\bM=[4, 8 ,4]$, $\bK=[4,6,8]$, and $N=25$ with $\bM=[4, 4, 4, 4, 4, 4, 4, 4, 4, 4, 4]$, $\bK=[4, 6, 8, 10, 12, 14, 15, 17, 19, 21, 23]$.}
	\label{fig:zoom_10_25}
\end{figure}

\begin{figure}[H]
	\centering
	\includegraphics[scale=0.5]{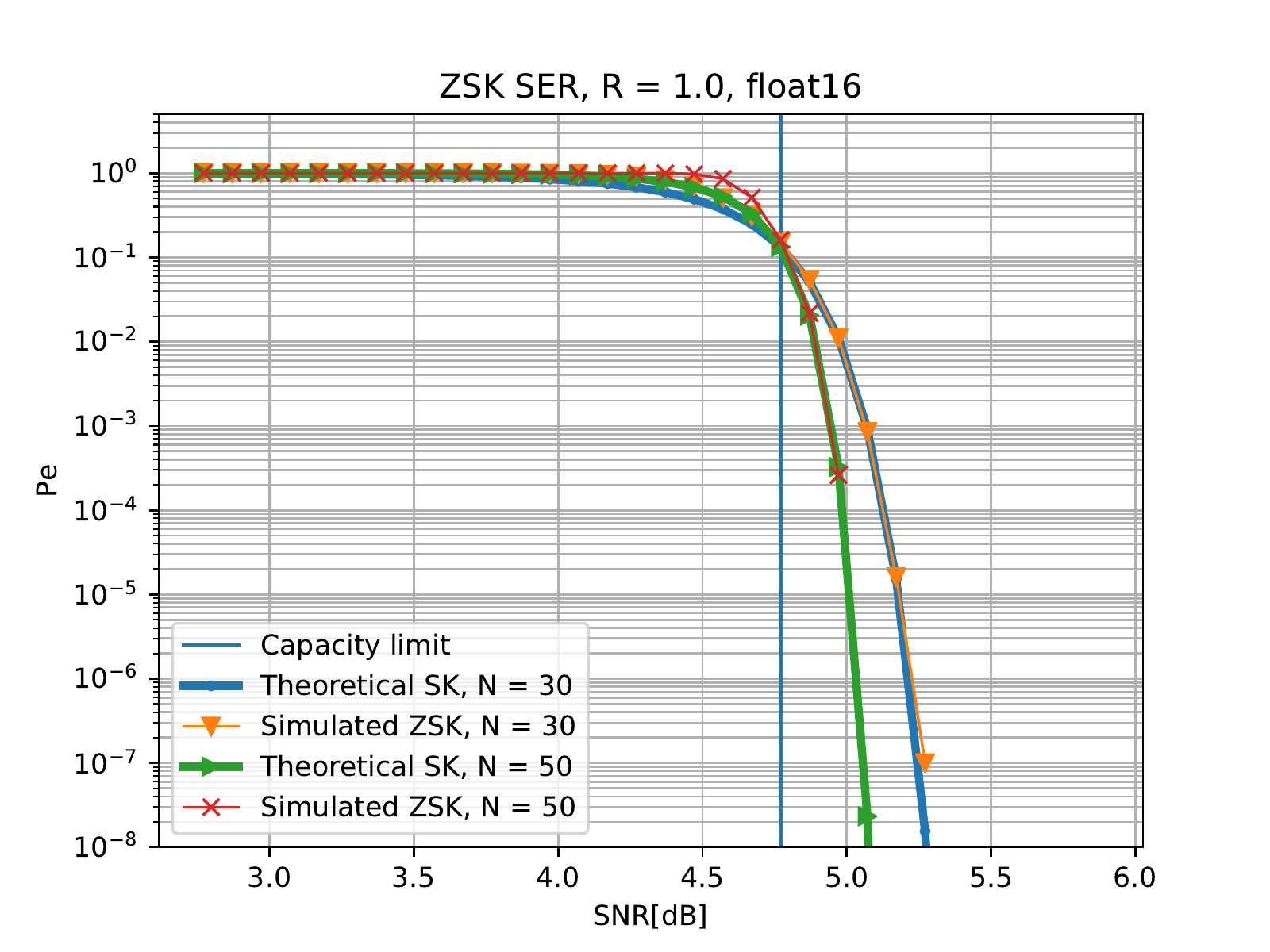}
	\caption{Zoom scheme symbol error rate for $N=30$ with $\bM=[4, 4, 4, 4, 4, 4, 4, 4, 4, 4, 4, 4, 4, 4]$, $\bK=[4, 6, 8, 10, 12, 14, 16, 18, 19, 21, 23, 25, 27]$ and $N=50$ with $\bM=[4, 4, 4, 4, 4, 4, 4, 4, 4, 4, 4, 4, 4, 4, 4, 4, 4, 4, 4, 4, 4, 4]$, $\bK=[5, 6, 8, 10, 12, 14, 16, 18, 20, 22, 24, 26, 28, 29, 31, 33, 35, 37, 39, 41, 43, 45, 47]$.}
	\label{fig:zoom_30_50}
\end{figure}

In the following we present some of the results achieved by running the proposed zoom scheme in a simulation, compared to the regular SK scheme. In Figure \ref{fig:float16_SK_no_zoom} we can see how the regular SK scheme fails even at a relatively small number of iterations, $N=10$, when using Float16. In Figure \ref{fig:float32_SK_no_zoom} we can see that even when using Float32, the SK scheme fails at a moderate number of iterations, $N=30$. In Figures \ref{fig:zoom_10_25} and \ref{fig:zoom_30_50}, we see that while regular SK fails after approximately 10 iterations, when using Float16, we can continue running our zoom scheme even up to $50$ iterations (or any other desired number of iterations). The capacity gap for $N=50$ is approximately $0.2\dB$ at $P_e=10^{-6}$.
Instead of storing the constellation size, $M$, we stored its logarithm $\log_2 M = NR$ in a short integer. The transmitted message, $i$, was stored as a long integer. The decoded message, $\widehat{i}$, was stored in the array of short integers, $i_0, i_1, \ldots, i_r$ such that $\widehat{i}=i_0+i_1+\ldots+i_r$. 
All the other variables were stored as Float16.
In the implementation of \eqref{eq:M_PAM} we create $\theta(i)$ in Float16 representation as a rounded value of the right hand side from the integers $i$ and $NR$. We implement \eqref{eq:a_from_i1} and \eqref{eq:PAM_i1_M1} similarly.   

\section*{Acknowledgment}
The authors would like to thank Ofer Shayevitz and Assaf Ben-Yishai for suggesting the use of a coupling argument in order to simplify the proofs used in the first version of this paper.



\end{document}